\documentclass{IOS-Book-Article} 
\usepackage{graphicx}
\usepackage{mathptmx}
\usepackage[utf8x]{inputenc}

\begin{document}

\pagestyle{headings}
\def\thepage{}

\begin{frontmatter}              

\title{Use and adaptation of open source software for capacity building to strengthen health research in low- and middle-income countries}


\author[A]{\fnms{Stefan} \snm{Hochwarter}%
\thanks{Corresponding Author: Stefan Hochwarter, E-mail: stefan@hochwarter.org}},
\author[A,B]{\fnms{Salla} \snm{Atkins}}
\author[A]{\fnms{Vinod K.} \snm{Diwan}}
and
\author[C,D]{\fnms{Nabil} \snm{Zary}}

\address[A]{Department of Public Health Sciences, Global Health (IHCAR), Karolinska Institutet, Stockholm, Sweden}
\address[B]{School of Health Sciences, University of Tampere, Finland}
\address[C]{Department of Learning, Informatics, Management and Ethics, Karolinska Institutet, Stockholm, Sweden}
\address[D]{Lee Kong Chian School of Medicine, Nanyang Technological University, Singapore, Singapore}

\begin{abstract}
Health research capacity strengthening is of importance to reach health goals. The ARCADE projects' aim was to strengthen health research across Africa and Asia using innovative educational technologies. In the four years of the EU funded projects, challenges also of technical nature were identified. This article reports on a study conducted within the ARCADE projects. The study focused on addressing challenges of video conferencing in resource constrained settings and was conducted using action research. As a result, a plugin for the open source video conferencing system minisip was implemented and evaluated. The study showed that both the audio and video streams could be improved by the introduced plugin, which addressed one technical challenge.
\end{abstract}

\end{frontmatter}
\pagestyle{empty}

\section{Introduction}

Efforts to strengthen health research capacity in low- and middle-income countries are needed\cite{r1}. The ARCADE HSSR and RSDH (African/Asian Research Capacity Development for Health Systems and Services Research/Social Determinants of Health) projects were two European Union-funded projects implemented from 2011 to 2015 and coordinated by the Division of Global Health at Karolinska Institutet in Stockholm. The projects aimed to strengthen health research across Africa and Asia by using innovational educational technologies\cite{r2}. The ARCADE projects can be divided into the following four interlinked components\cite{r3}. The first component of the ARCADE projects was the development and delivery of online courses on global health topics.

E-learning can be divided into synchronous and asynchronous education or e-learning. The latter one, asynchronous education, provides education from the teacher to the student even if both are not online at the same time. Thus, in simple terms asynchronous education uses technologies and tools like e-mail, panels or e-learning platforms like Moodle. The strengths of asynchronous education lies within its flexibility on both time and place and its low requirements on the bandwidth, which is an important point for resource constrained settings\cite{r4}. 
On the other hand, asynchronous e-learning has also weaknesses, such as no possibility for immediate feedback, discussions on boards will last much longer on complex topics compared to face-to-face discussions and the lack of social interaction which may result in students not feeling connected to each other. Moreover, this method requires a severe discipline from the students as they need to manage their time of learning on their own\cite{r5}. To overcome the disadvantages of e-learning and to take into account bandwidth challenges, the ARCADE projects used mostly blended learning which is a combination of synchronous (face-to-face online or in person) and asynchronous (e-learning) methods. 

The ARCADE project used different open source software to prepare, organise and deliver e-learning courses. Almost 55 courses were developed and delivered to over 920 postgraduate students in Africa, Asia and Europe using e-learning principles and specifically blended learning\cite{r6}. The open source e-learning platform Moodle\footnote{https://moodle.com} was the main entry point for the students. Synchronous distance education was delivered with the help of minisip\footnote{https://github.com/csd/minisip}, which is an open source Session Initiation Protocol (SIP) implementation developed by ARCADE's project partner Royal Institute of Technology Stockholm\cite{r7}. However, as this partnership weakened over time, alternatives to minisip were evaluated and used. Content management was implemented with the help of Alfresco Community Edition\footnote{https://www.alfresco.com/de/products/enterprise-content-management/community}, an Enterprise Content Management platform. Dissemination and publication of research findings were presented online at the self-hosted project's web site (www.arcade-project.org) which is powered by Wordpress\footnote{https://wordpress.org/}.  

Färnman et. al. investigated the challenges which the ARCADE project team came across during the four years of project runtime by interviewing 16 participants from 12 partner institutions. The main challenges for e-learning included problems in technology, availability of skilled technical staff across implementation sites and attracting students' interest in courses. The report also points out the high demand on bandwidth and software deficiencies in resource restrained settings\cite{r3}. This is specifically true for synchronous distance education. The limitation of the bandwidth, poor image and video quality as well as connectivity issues are challenges when video teleconferencing systems are used\cite{r8}. In this article we will demonstrate how open source software can be used to overcome bandwidth limitations.

\section{Methods}

The aim of this study was to evaluate how Open Source Software (OSS) can be used and optimised for for distance education, particularly in the area of health research/education in a global setting. As the study took place within the ARCADE project which used the minisip software, the underlying study question can be stated as followed.

\begin{itemize}
  \item How can OSS such as minisip improve the delivery of synchronous health education in resource-restricted environments?
\end{itemize}

The design of the study was defined by an action research framework\cite{r9}. Both qualitative and quantitative data were used, hence we followed a mixed research approach. This study can be divided in the different phases of action research.

\begin{enumerate}
 \item \textbf{Observe} A questionnaire was conducted with the aim to explore the ARCADE RSDH's teaching activities and its priorities to the system and quantitative data was collected at St. John's Research Institute in India.
 \item \textbf{Reflect} The data from the questionnaire and measurement were analyzed and possible improvements were identified.
 \item \textbf{Act} Based on the reflections, an improvement was chosen for implementation.
 \item \textbf{Evaluate} The implemented improvement was tested and evaluated at Wuhan's Tongji Medical College.
\end{enumerate}

The questionnaire consisted of one multiple choice question and three free-text questions and was answered by ARCADE partners from Sweden (Karolinska Institutet) and India (St. John's Research Institute). A quantitative measurement of the Minisip performance using Wireshark was done at St. John's Research Institute in Bangalore.

The Minisip performance for the ARCADE RSDH project was evaluated and improvements were suggested, implemented and tested. The study took place at ARCADE partners in Sweden (Karolinska Institutet), India (St. John's Research Institute Bangalore) and China (Wuhan's Tongji Medical College). During the study, the focus of the terms "adaptability" and "performance" were narrowed down based on the outcome of the questionnaire. 

Data were analysed using the statistical environment of R\footnote{https://www.r-project.org/}. The R package RQDA\footnote{http://rqda.r-forge.r-project.org} was used to organise and analyse qualitative data. Captured traffic from minisip was recorded and filtered by using Wireshark\footnote{https://www.wireshark.org/}. Wireshark offers detailed analyse function for SIP calls and Real-Time Transport Protocol (RTP) streams, including package loss and jitter. 

\section{Results}

In the first step (Phase "Observe"), the background, aims and requirement of the underlying project was investigated by a questionnaire. A quantitative analysis of minisip at St. John's Research Institute in Bangalore showed that one challenge was the rather high Packet Delay Variation (PDV) or so-called "jitter". Based on the questionnaire in the first step and the measurement outcome (Phase "Reflect"), we decided to implement a plugin for minisip with the aim to minimise the PDV. Minisip offers an interface to insert an extension at different stages of the traffic flow. The implemented extensions were inserted after the RTP pipeline at the project partner at Wuhan's Tongji Medical College, China (Phase "Act").

Two different algorithms were implemented to improve the traffic flow. As the audio packets all had the same size, a simple leaky bucket algorithm was sufficient. However, the video packets were of different size and therefore the byte-based token bucket algorithm was chosen\cite{r10}.

\begin{figure}
  \centering
  \includegraphics[width=0.6\textwidth]{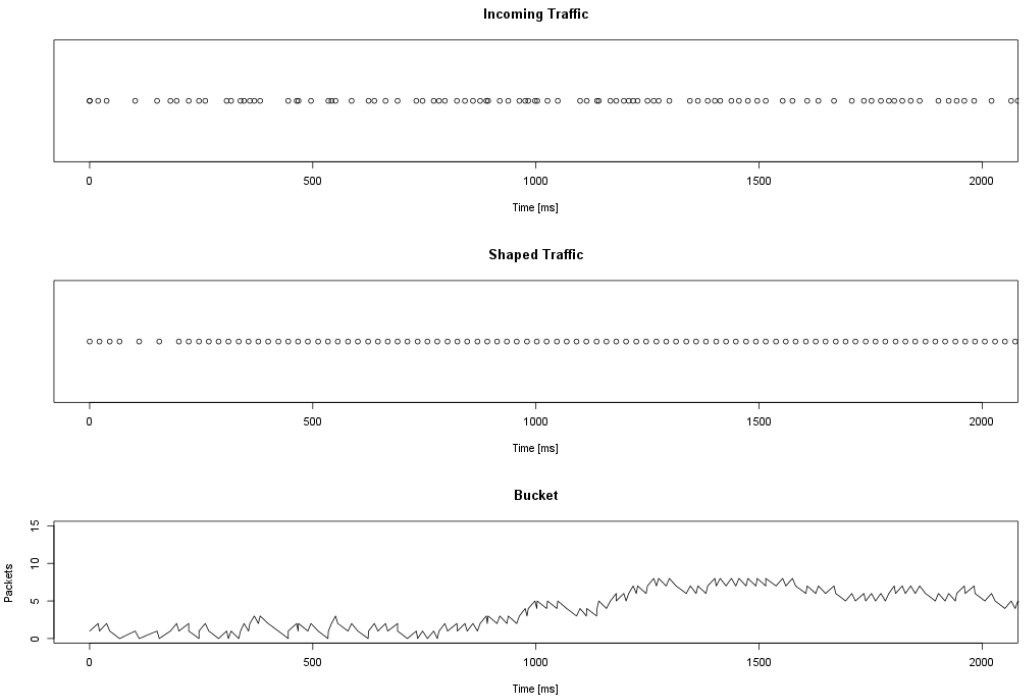}
  \caption{Measurements from the leaky bucket plugin for the audio stream (Wuhan, China)}
\end{figure}

The final phase of the action research was to evaluate the outcome of the implementation. Figure 1 shows the output of the leaky bucket extension for incoming audio. The first diagram visualises the incoming packets, each dot represents one incoming packet (each 125 Bytes) before the extension is shaping the traffic. The second diagram shows the manipulated traffic, again one dot represents one packet. Finally, the third diagram represents the current content of the bucket with a maximum capacity of 15 packets.

\begin{figure}
  \centering
  \includegraphics[width=0.6\textwidth]{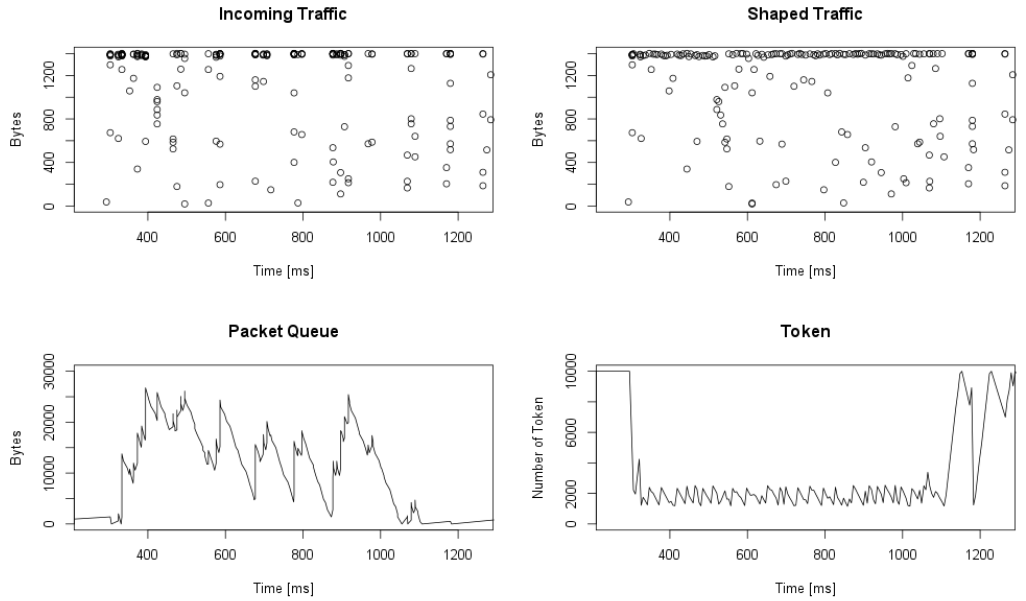}
  \caption{Measurements from the leaky bucket plugin for the video (H.264) stream (Wuhan, China)}
\end{figure}

Figure 2 illustrates the effect of the token bucket algorithm. The diagram "incoming traffic" visualises the incoming video packets over time and the diagram "shaped traffic" shows the resulted traffic after the token bucket extension. One dot represents one packet and the vertical axes shows the packet size. Two important parameters of the extension are also visualised: The packet queue shows the number of bytes (not packets)  waiting for transmission and the last diagram represents the token available for traffic. One token is equal to one Byte in the used configuration. 

As shown in Figure 1 and Figure 2, the PDV was successfully minimised for both audio and video streams.  

\section{Discussion}

The study demonstrated with a specific example, how open source software can be adapted for capacity building in low and middle income countries based on a scientific background. However, to address the challenges which the ARCADE project team encountered during their four year project runtime, different strategies should be used. For example, investment in IT infrastructure and educating technical staff would be two possible strategies.

During this study we also faced a typical risk when using a open source software that is not backed by a strong community. The main developer of minisip stopped his contribution and since then minisip is not further developed. The initial strength, namely the close partnership with the developer and thereby the possibility to receive an adapted software solution for our settings, turned out to be also a high risk factor.

Nevertheless, we believe that the use of open source software has high potential and advantages, especially in resource restrained settings. Within a limited time-frame, the project team successfully introduced an improvement for a specific problem based on measurements and interviews. This was made possible through an open source software that had a clean codebase and extensive documentation. Therefore external developers and contributors were enabled to introduce changes and improvements. As part of a future study, the plugins could also be integrated and tested in other SIP implementations. Finally, the implemented plugins could be further improved by automatically detecting the algorithms' parameters.


\begin{thebibliography}{99}

\bibitem{r1}
McKee, M., Stuckler, D., \& Basu, S. (2012). \textit{Where there is no health research: what can be done to fill the global gaps in health research?} PLoS Med, 9(4), e1001209.

\bibitem{r2}
Protsiv, M., Rosales-Klintz, S., Bwanga, F., Zwarenstein, M., \& Atkins, S. (2016). \textit{Blended learning across universities in a South–North–South collaboration: a case study.} Health Research Policy and Systems, 14(1), 67.

\bibitem{r3}
Färnman, R., Diwan, V., Zwarenstein, M., \& Atkins, S. (2016). \textit{Successes and challenges of north–south partnerships–key lessons from the African/Asian Regional Capacity Development projects} Global Health Action, 9.

\bibitem{r4}
Branon RF, Essex C. \textit{Synchronous and asynchronous communication tools in distance education.} TechTrends. 2001;45(1):36-36

\bibitem{r5}
Hrastisnski S. \textit{Asynchronous and synchronous e-learning.} Educause quarterly. 2008;31(4):51-55

\bibitem{r6}
Atkins, S., Marsden, S., Diwan, V., \& Zwarenstein, M. (2016). \textit{North–south collaboration and capacity development in global health research in low-and middle-income countries–the ARCADE projects} Global Health Action, 9.

\bibitem{r7}
Rosenberg J, Schulzrinne H, Camarillo G, Johnston A, Peterson J, Sparks R, et al. \textit{SIP: session initiation protocol.} RFC 3261, Internet Engineering Task Force; 2002. Available from: http://www.ietf.org/rfc/rfc3261.txt.

\bibitem{r8}
Frehywot S, Vovides Y, Talib Z, Mikhail N, Ross H, Wohltjen H, et al. \textit{E-learning in medical education in resource constrained low-and middle-income countries.} Human resources for health. 2013;11(1):1–15.

\bibitem{r9}
 Reason P, Bradbury H. \textit{The SAGE Handbook of Action Research: Participative Inquiry and Practice. vol. 2nd} Reason P, Bradbury H, editors. SAGE; 2008.

\bibitem{r10}
Tanenbaum A. \textit{Computer Networks} 4th ed. Prentice Hall Professional Technical Reference; 2002.

\end{thebibliography}
\end{document}